\documentclass[superscriptaddress, preprint]{revtex4-2}
\usepackage{graphicx}
\graphicspath{.}

\usepackage{amssymb}
\usepackage{amsmath}
\usepackage{amsthm}
\usepackage{amsfonts}
\usepackage{xfrac}
\usepackage{wasysym}
\usepackage{textcomp}
\usepackage{bbm}

\usepackage{braket}
\usepackage[version = 4]{mhchem}
\usepackage{epstopdf} 

\usepackage{hyperref}
\usepackage{array}
\usepackage{xcolor}
\usepackage{soul}

\begin{document}

\author{Stefano Marin}
\email{stmarin@umich.edu}
\affiliation{Department of Nuclear Engineering and Radiological Sciences, University of Michigan, Ann Arbor, MI 48109, USA}

\author{Ivan A. Tolstukhin}
\email{itolstukhin@anl.gov}
\affiliation{Physics Division, Argonne National Laboratory, Lemont, IL 60439, USA}

\author{Nathan P. Giha}
\affiliation{Department of Nuclear Engineering and Radiological Sciences, University of Michigan, Ann Arbor, MI 48109, USA}

\affiliation{Physics Division, Argonne National Laboratory, Lemont, IL 60439, USA}

\author{Fredrik Tovesson}
\affiliation{Physics Division, Argonne National Laboratory, Lemont, IL 60439, USA}

\author{Vladimir Protopopescu}
\affiliation{Oak Ridge National Laboratory, Oak Ridge, TN 37830, USA}

\author{Sara A. Pozzi}
\affiliation{Department of Nuclear Engineering and Radiological Sciences, University of Michigan, Ann Arbor, MI 48109, USA}
\affiliation{Department of Physics, University of Michigan, Ann Arbor, MI 48109, USA}

\title{Measurement of fragment-correlated $\gamma$-ray emission from $^{252}$Cf(sf)}
\date{\today}

\begin{abstract}
This paper presents recent experimental results on the yield of prompt fission $\gamma$ rays from the spontaneous fission of $^{252}$Cf. We use an ionization chamber to tag fission events and measure the masses and kinetic energies of the fission fragments and trans-stilbene organic scintillators to measure the neutrons and $\gamma$ rays emitted by the fission fragments. The combination of the ionization chamber and trans-stilbene scintillators allows us to determine the properties of neutrons and $\gamma$ rays in coincidence with the fragments. The yield of $\gamma$ rays is known to be influenced by the angular momenta (AM) of the fission fragments. We present new experimental evidence indicating that the total $\gamma$-ray multiplicity, \textit{i.e.}, the sum of both fragments' emission, saturates at sufficiently high internal fragment excitation energies. We also observe distinct behaviors for the yield of $\gamma$ rays from the light and heavy fragment, which for certain mass and total kinetic energy (TKE) regions are weakly or anti-correlated, indicating the presence of complex AM generation modes. We also observed a mass- and TKE-dependent anisotropy of the $\gamma$ rays, which challenges and expands on the conventional notion that the fragments' AM are always aligned perpendicularly to the fission axis. Moreover, the dependence of the anisotropy on mass and TKE indicates a dependence of these properties on the specific fission channels, thus providing an insight into the deformations and dynamics in fission and their connection with experimentally observable quantities. 
\end{abstract}


\maketitle

\section{Introduction} 
\label{sec:intro}

Nuclear fission has received renewed interest in the past decade, thanks in part to new experimental data~\cite{Gook2014, Travar2021, Wilson2021} and large computational models of nuclei~\cite{Bulgac2021, Schunck2022}. However, there are several aspects of nuclear fission that still remain unresolved, in particular, the generation of the fragment angular momenta (AM) in fission. AM is an important aspect to understand in fission because of its role in the description of deformation processes, of which fission is a chief and extreme example~\cite{BohrBook}. At the same time, AM strongly influences the feeding of rotational bands in fission fragments and is thus important in the measurement and spectroscopy of $\gamma$ rays. 

The main unresolved question revolves around identifying the mechanism responsible for generating AM in the fission fragments, particularly considering that the fissioning nucleus possesses minimal or no initial AM. A current obstacle in answering this question is that several conceptually distinct theoretical models have been proposed for the generation of AM, all of which have reasonable but quite distinct assumptions and predictions. Indeed, at least three mechanisms for the generation of fragment AM have been proposed: deformation of fission fragments~\cite{Marevic2021, Bulgac2021}, statistical excitation of rotational modes~\cite{Moretto1989, Randrup2021, Gonnenwein2005}, and Coulomb torque~\cite{Hoffman1964, Scamps2022}. The predictions made by these models tend to agree on established results, such as the average magnitude of the fragment AM, but disagree on other more complex observables, such as the relative orientations of the fragments' AM. The purpose of the current experimental work is to add to the rich literature of fission observables data, enabling the discrimination of the theoretical models and potentially elucidating the mechanism responsible for AM generation in fission.

In an experimental setting, it is challenging to gain direct access to the fragment AM. This quantity is related to a rotational symmetry of the nuclear wave function of the nucleus that decays quickly by neutron and $\gamma$-ray radiation. Some of the most common and successful techniques that have been used in determining the AM are $\gamma$-ray spectra~\cite{Wilson2021}, isomeric yield ratios~\cite{Rakapoulos2018, Chebboubi2017}, $\gamma$-ray angular distributions~\cite{Wilhelmy1972, Hoffman1964}, and $\gamma$-ray multiplicity distributions~\cite{Nifenecker1972, SchmidFabian1988}. The work presented in this paper draws conclusions on the fragment AM based on the multiplicity and angular distribution of the $\gamma$ rays they emit.

In Section~\ref{sec:inst} we briefly discuss the specifics of the experimental system. The study of $\gamma$-ray multiplicity for different fragment mass regions and kinetic energies is presented in Section~\ref{sec:multGam}. In Section~\ref{sec:angGam} we study the angular distribution of $\gamma$ rays with respect to the fission axis, attempting to determine the polarization of the fragment AM in fission. We discuss the results of this experiment and their interpretation in Section~\ref{sec:conc}.

\section{Experimental Methods}
\label{sec:inst}

\subsection{Instruments}

The experiments we present in this paper use the connection between the $\gamma$-ray yield and the fission fragments' AM. In particular, just as neutron multiplicity is often used as a proxy observable for the fragment excitation energy ~\cite{Nifenecker1972, Brosa1990}, $\gamma$-ray multiplicity and angular distribution carry important information regarding the fragments' AM. By correlating these $\gamma$-ray observables with coincident data on fragment properties, such as masses and TKE, we can study trends in AM as the underlying fragment properties change.  

The experiment was performed using the FS-3 array of trans-stilbene organic scintillators coupled to the Argonne National Laboratory twin Frisch-gridded ionization chamber (TFGIC). A recent paper~\cite{Marin2023} details the technical aspects and performance of this experimental setup.

The target in this experiment was prepared at Oregon State University and consists of a molecular plating of $^{252}$Cf on a $\approx 100$ $\mu \text{g} \ \text{cm}^{-2}$ carbon foil. The thin target allows both fragments to escape into the separate sides of the chamber and be detected. The $2E$ method~\cite{Budtz1987} is used to determine the fragment masses from the measured kinetic energies. The molecular plating technique leaves a non-uniform ``crud'' of cracked solvent molecules, which makes the resolution of the system slightly inferior to those using more uniform types of depositions. We have accounted for this effect in our analysis; however, remnants of the correction are still noticeable at larger angles, where the fragments have to travel through a greater amount of material. For this reason, we have constrained the analysis of the angular distribution only to $| \cos \theta | > 0.3$, where $\theta$ is the angle between the fragment direction and the axis of the fission chamber, which is itself normal to the source plane.

Two configurations were used to provide the results of the work presented here: one including all detectors of the FS-3 system and a $^{252}$Cf  target with total activity of approximately $9$ kBq, and one where only the two FS-3 detectors collinear with the TFGIC are used. For the second configuration, a stronger $^{252}$Cf  source of total activity $\sim 130$ kBq is used to compensate for the reduced efficiency of the detectors.

\subsection{Multiplicity and spectrum analysis}

The full FS-3 detector array consists of 39 active detectors placed in spherical configuration around the source, and a geometric coverage of approximately $12 $\%. Standard unfolding techniques, based on the simulated energy-dependent response matrix of the system to $\gamma$-rays, has been applied to recover the multiplicity and spectra of the measured $\gamma$ rays. Given the  broad energy resolution of trans-stilbene detectors,  peaks cannot be resolved but gross features of the spectrum can be seen. 

\begin{figure}[!htb]
\centering
\includegraphics[width=0.8 \linewidth]{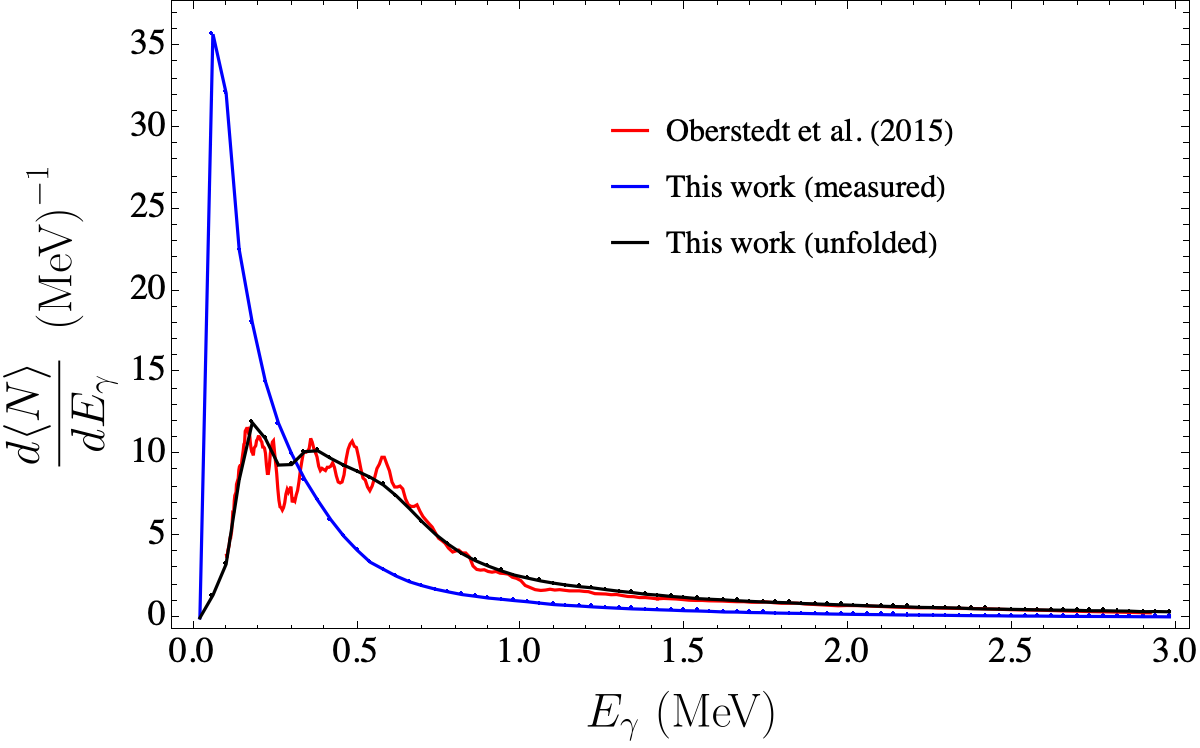}
\caption[Unfolded spectrum]{The measured $\gamma$-ray spectrum (scaled here simply by the inverse of the simulated efficiency) is unfolded using simulated $\gamma$-ray response matrices. The resulting unfolded spectrum reproduces the shape of the emitted $\gamma$-ray spectrum as observed by recent experiments~\cite{Oberstedt2015}. Since the size of the statistical error bars is smaller than the line width, the error bars are not included. Systematic errors, associated with both the measurement apparatus and the unfolding process, are not included either.}
\label{fig:specUnf}
\end{figure}

Fig.~\ref{fig:specUnf} shows the measured and unfolded spectra of $\gamma$ rays detected during this experiment. We have compared our unfolded results to a recent measurement by Oberstedt \textit{et al.}~\cite{Oberstedt2015}, which used dedicated inorganic scintillators. While the finer features are not exactly reproduced, we can see from the figure that the unfolding procedure of the $\gamma$-ray spectrum significantly improves our measurements and allows us to validate the simulated experimental response. 

In addition to the total $\gamma$-ray multiplicity, it is also possible to extract from the experiment the individual $\gamma$-ray multiplicities emitted from each fragment. This is performed by applying the Meier-Leibniz Doppler-shift method~\cite{Travar2021}. In this method, only detectors collinear with the TFGIC are used, \textit{i.e.}, the second configuration of our experiment, and the aberration of $\gamma$-rays yield is used to infer the average multiplicity emitted by each fragment. 

\subsection{Angular distribution and correlation analysis}

\begin{figure}[!htb]
\centering
\includegraphics[width=0.8 \linewidth]{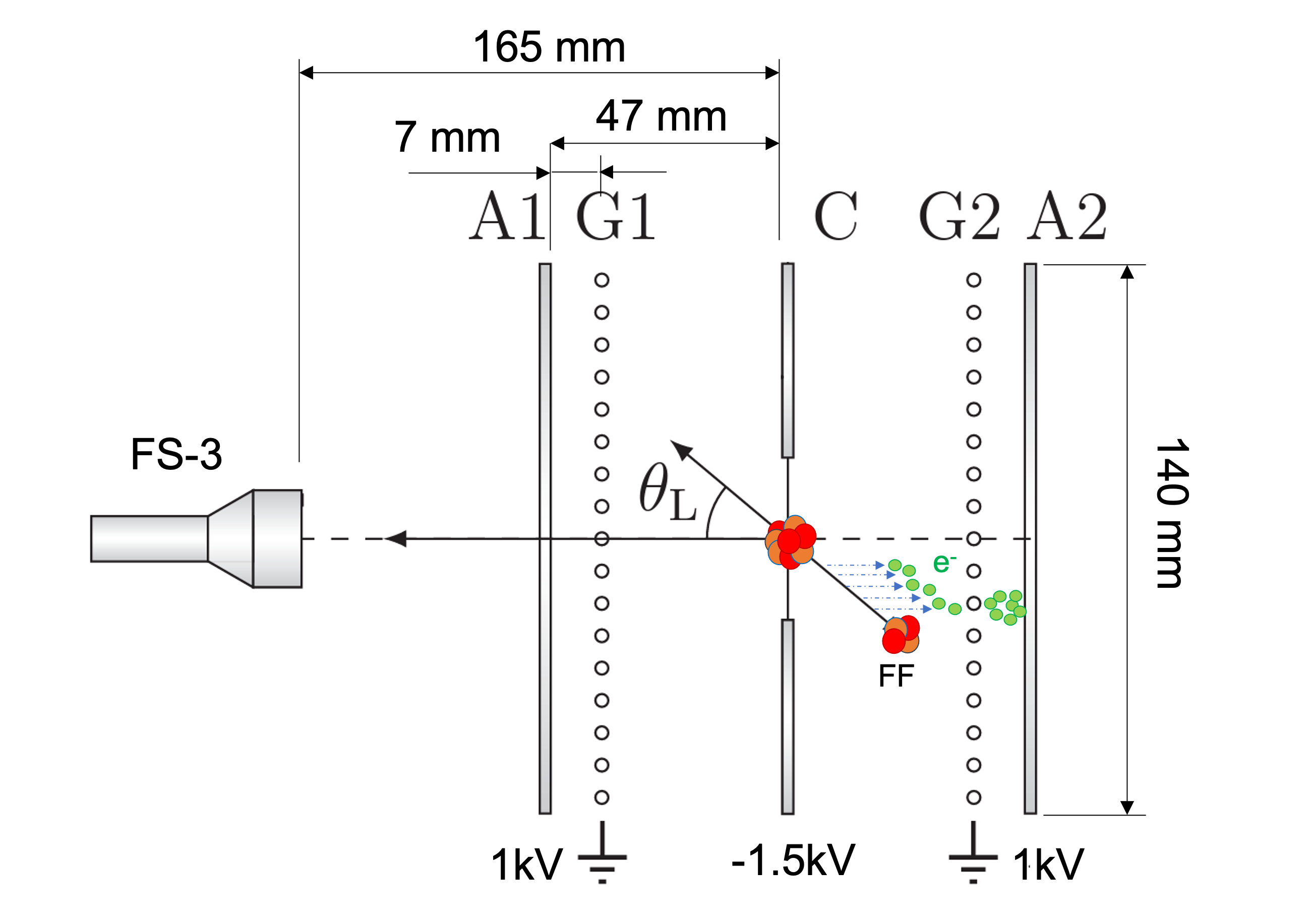}
\caption[Angular distribution technique]{Schematic representation of the present experiment, adapting a figure first used in Ref.~\cite{Gook2014}. The geometric and electric specifications of our system are indicated on the figure. The angle between the chamber axis and the fission axis, taken to be the direction of the light fragment, is indicated as $\theta_L$. }
\label{fig:angTech}
\end{figure}

The study of angular distributions and correlations of radiation is a well-established technique in all branches of nuclear and atomic physics, especially when dealing with AM~\cite{Wilhelmy1972}. In this experiment, we measure the angular distribution of $\gamma$ radiation using the technique developed by G\"{o}\"{o}k \textit{et al.}~\cite{Gook2014}, who measured the angular distribution of neutrons from fission. In particular, in Fig.~\ref{fig:angTech} we adapt Fig. 1 from their paper to show the relevant dimensions.

In spontaneous fission it is customary to indicate the axis of fragment motion as the fission axis; the direction of the light fragment is here taken by convention to be the direction of the fission-axis vector. All angular variables are expressed with respect to this direction. In Fig.~\ref{fig:angTech}, the angle indicated as $\theta_L$ is the angle that the light fragment makes with the axis of cylindrical symmetry of the TFGIC. The angular distribution of $\gamma$ rays is determined by measuring the count rate in the detector aligned with the TFGIC axis. This technique allows us to use the same detector for all measurements, thereby simplifying the analysis and corrections applied to the data. Furthermore, this type of angular measurement can be employed even when an azimuthal measurement of the fragment direction is not available, as is the case in this experiment. 

\section{Multiplicity}
\label{sec:multGam}

 In this section, the focus turns to the development of results concerning the multiplicity of $\gamma$ rays from fission. We start by presenting results related to neutron multiplicity, which serves to validate the detection system, considering the extensive study of neutrons in the literature. Following that, an analysis is conducted on the total multiplicity of $\gamma$ rays with respect to both fragment masses and TKE. Finally, the Meier-Leibniz Doppler-Shift method is applied to the data collected in the second configuration, utilizing only two detectors. In this third step, the $\gamma$-ray multiplicity emitted by each individual fragment is presented.

\subsection{Neutron multiplicity}

\begin{figure}[!htb]
\centering
\includegraphics[width=0.8 \linewidth]{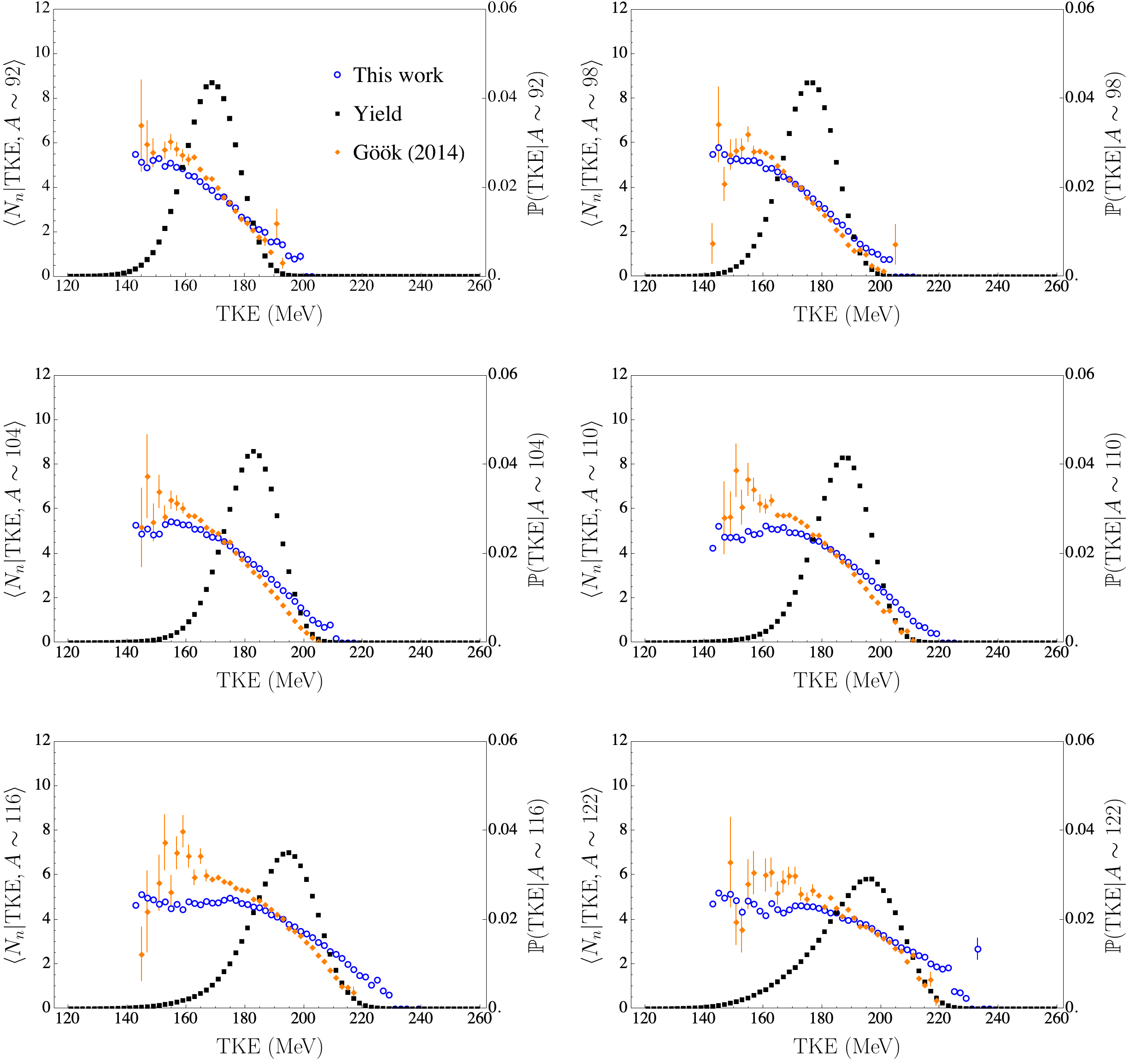}
\caption[Neutron multiplicity]{Measured mass-binned TKE-dependent neutron multiplicity compared to studies performed by G\"o\"ok \textit{et al.}~\cite{Gook2014}. The yield of TKE, at the specified mass, is shown as the black curve, with arbitrary scaling.}
\label{fig:gridNeut}
\end{figure}

The multiplicity of fission neutrons is one of the most intensely studied topics in the field, due in large part to its importance in technological applications of fission. The dependence of the emitted neutron multiplicity on the fragment TKE for selected masses is shown in Fig.~\ref{fig:gridNeut} and compared to the results by G\"o\"ok \textit{et al.}~\cite{Gook2014}. In contrast to the $\gamma$-ray multiplicity, which is obtained by unfolding the measured multiplicity using a simulated matrix response, the emitted neutron multiplicity is determined by a standard efficiency scaling, where the efficiency has been simulated and the resulting mean neutron multiplicity agrees well with the known value of $\langle N_n \rangle \approx 3.75$.  We observe the clear anti-correlation between neutron multiplicity and TKE. This correlation aligns with expectations, as neutron multiplicity is known to be an excellent information carrier for the total fragment excitation energy, $E^*$, which is the sum of the two fragments' excitation energies. This quantity is related to the total energy released in fission, $Q$, and TKE through the equation
\begin{equation}
    E^* = Q(A) - TKE \ .
\label{eq:estar}
\end{equation}
The total energy released has a mass dependence, based on the different binding energies of the fission fragments. The $Q$ value of the reaction depends on the charges of the reaction products as well as their masses. We average the $Q$ value over the tabulated distribution of charges for each isobar. While the charge distribution is quite narrow, this averaging procedure introduces fluctuations of the order of $3-4$ MeV. Thus, we expect the energy resolution in $E^*$ to be of a similar magnitude. The x-intercept of the neutron multiplicity on TKE is thus strongly related to the total $Q$ value of that fission reaction, although it is an underestimate, as a minimum amount of energy is required to emit the first neutron. The diminishing slope of neutron multiplicity with decreasing TKE can be attributed to several factors. First, nucleons in the fragment become more tightly bound as neutrons are emitted, thus increasing the energy needed to remove them. Second, we cannot exclude that the $E^*$ generated during fission is not all available for neutron emission, and might be found in nuclear degrees of freedom, such as rotational motion and intrinsic deformations, that do not contribute to the neutron evaporation process.

Our data compares well with the data from G\"o\"ok \textit{et al.}, indicating that our detection system is capable of properly identifying trends in the multiplicity of the emitted radiation and the underlying fragment. We note that especially for rarer fragmentations of extreme asymmetry and symmetry, the data are significantly broadened. The degradation of resolution, which we attribute to a combination of intrinsic chamber resolution and energy broadening in the target itself, leads to less pronounced features and flatter slopes in both energy-dependent and mass-dependent quantities.

\subsection{Total $\gamma$-ray multiplicity}

The total $\gamma$-ray multiplicity accompanying a fission event encodes information regarding the total fragments' AM. In fact, after the emission of neutron and statistical $\gamma$ rays, the fragments' rotational energy is generally dissipated by emission of collective $\gamma$ rays along rotational bands. The most common type of emission is electric quadrupole (E2) which lowers the fragment AM by $2$$\hbar$~\cite{Serot2010}. In an ideal scenario, it would be possible to establish a direct correlation between $\gamma$-ray multiplicity and AM through a proportional relationship. However, this overly simplifies picture of the de-excitation process is somewhat misleading; while correlations exist between $\gamma$-ray multiplicity and AM, their relationship is not as simple. A significant portion of the $\gamma$ rays, estimated with model calculations~\cite{Litaize2015} to be as many as half of all $\gamma$ rays on average, are emitted from states in the quasi-continuum, and predominantly favor electric dipole (E1) transitions. The AM dissipated in each of these statistical transitions is not known, and is a topic of current research~\cite{Stetcu2021, Marin2022}. Furthermore, within the collective $\gamma$-ray transitions, we can find different $\gamma$-ray multipolarities, such as magnetic dipole (M1), as well as E2 emissions that are not stretched, \textit{i.e.}, that connect states which have a difference of AM smaller than $2 \hbar$. 

With the above disclaimers, there is still compelling value in studying the total $\gamma$-ray multiplicity in the absence of more refined measurements. We can study the trends of $\gamma$-ray multiplicity with respect to fragment variables, thus assuming that the aforementioned effects do not themselves vary significantly with fragment properties. This assumption is more valid if we focus on regions of large $E^*$, where statistical $\gamma$-ray emission approximately saturates, and changes in $\gamma$-ray yield can be attributed primarily to rotational-band feeding. 

\begin{figure}[!htb]
\centering
\includegraphics[width=0.8 \linewidth]{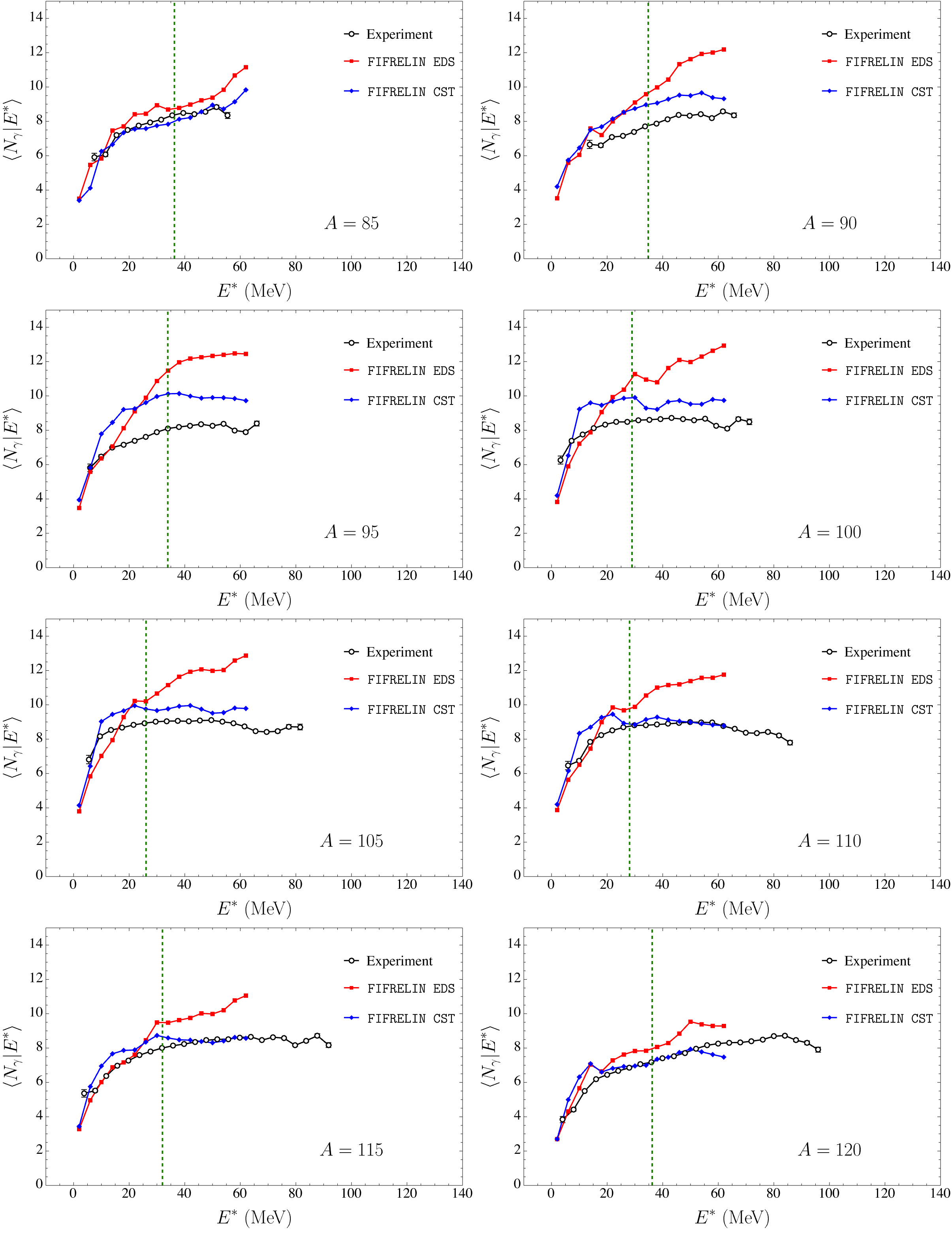}
\caption[Gamma multiplicity]{Comparison of measured total $\gamma$-ray multiplicity, summed contributions from each fragment, with two different model calculations performed in \texttt{FIFRELIN}. Each panel corresponds to bins centered at the specified light fragment mass $A$, and with a bin width of $5$ AMU. The dashed vertical line corresponds to the average total $E^*$ for the specified fragmentation.}
\label{fig:gridExcCompFire}
\end{figure}

Fig.~\ref{fig:gridExcCompFire} shows the trends of total $\gamma$-ray multiplicity with respect to $E^*$, as calculated using Eq.~\ref{eq:estar}. To highlight the role of AM, we compare the experimental data to calculations performed using \texttt{FIFRELIN}, which treats the excitation and de-excitation of fission fragments~\cite{Serot2010, Chebboubi2017}. Specifically, we compare our data to two different calculations, where in both cases the AM is sampled uniformly from all available states at the simulated fragment energy. A description of these two calculations and their differences appears in the recent Ref.~\cite{Piau2023}. The first calculation, labeled here \texttt{FIFRELIN CST}, uses a realistic model for the excitation energy dependence of the level density, but the angular momentum dependence is not considered. Specifically, above the energy at which complete level scheme exist, the density of states with a given AM is independent of $E^*$. Thus, at sufficiently high energy, generally on the order of a few MeV, the angular momentum becomes independent of the excitation energy. The second calculation, called \texttt{FIFRELIN EDS}, includes an energy dependence of the spin cut-off distribution, which takes into account the moment of inertia of the fragments. In this calculation, the density of states with higher AM increases with $E^*$, and so does the sampled distribution of fragments' AM. It should be noted that this second calculation is technically more accurate, as it is known that the AM distribution changes with $E^*$~\cite{BohrBook}. The investigation of Ref.~\cite{Piau2023} has also shown that the \texttt{FIFRELIN EDS} calculation, especially with the Hartree-Fock Bogoliubov level density treatment, reproduces the fragment AM sawtooth observed by Wilson \textit{et al.}~\cite{Wilson2021}.

The experimental data can be described as being made up of two regions: at the lowest excitation energies, the yield of $\gamma$ rays increases very quickly with $E^*$; at energies above $\approx$ 10 MeV the rate changes significantly, in some cases reaching a leveled-off plateau. The division between the two region likely originates from the onset of neutron emission, which dominates at high $E^*$. At low $E^*$ neutron emission is not energetically possible, so all the energy in the system goes into $\gamma$-ray emission. After both fragments acquire sufficient energy to emit neutrons, the yield of $\gamma$-rays should increase only if the fragment AM is also increasing with $E^*$, as a higher AM results in a rotational energy that is not easily depleted by neutron emission.

We see from our data that for very asymmetric, $A_L < 90$, and very symmetric, $A_L > 115$, mass splits, the yield of $\gamma$ rays continues to increase with $E^*$ even at higher energies, although at a much slower rate than in the first region. A different behavior is observed for more standard mass splits, where the yield of $\gamma$ rays plateaus and does not appear to increase significantly with $E^*$. The lack of a plateau at the most symmetric and asymmetric mass splits should not be taken as an indication of an increasing AM. 

Comparing our experimental results to the \texttt{FIFRELIN CST} calculations for these regions, we see that an increase in $\gamma$-ray yield can be observed at high $E^*$ even when the AM is independent of energy. That is because for these highly asymmetric and symmetric cases the energy partition between the fragment is far from even. Looking at a neutron sawtooth distribution, \textit{i.e.}, the mean neutron multiplicity as a function of fragment mass, we can see that in the case of high asymmetry the heavy fragment takes most of the $E^*$, whereas the situation is reversed for the symmetric case. This unequal partition of the total $E^*$ implies that the total energy has to be significantly higher for both fragments to have enough internal excitation energy to make neutron emission dominant in both fragments. 

Greater understanding can be derived by examining the more standard mass splits, in which the excitation energy is more evenly distributed. Consequently, a saturation of the rotational-energy independent gamma-ray emission occurs at lower excitation energies. As we see from Fig.~\ref{fig:gridExcCompFire}, it is in this mass region, $100 < A_L < 110$, that the two calculations are at variance with one another. In \texttt{FIFRELIN EDS}, excitation energy is coupled to AM and thus the yield of $\gamma$ rays keeps increasing with $E^*$. In \texttt{FIFRELIN CST}, on the other hand, after an initial increase the $\gamma$-ray yield remains relatively constant. The experimental data we collected aligns more closely with the behavior of \texttt{FIFRELIN CST}, although quantitatively the agreement is not excellent. 

The agreement of our data with the \texttt{FIFRELIN CST} calculations, see for example $A_L = 105$ and $110$ in Fig.~\ref{fig:gridExcCompFire}, is surprising from a physical perspective. The density of AM states does indeed depend on the excitation energy, so a statistical population of AM states should result in positive correlations between $E^*$ and AM. Since we observe that our data agrees better with the \texttt{FIFRELIN CST} calculations, this experiment indicates the existence of non-statistical mechanisms of AM generation. Of course, several other variables including neutron dissipation of AM should be controlled for in future experiments. 


We observe a drop in the $\gamma$-ray multiplicity at the highest $E^*$, or equivalently the lowest TKE for a given mass split. However, the statistics are quite poor in this region. Other systematic errors, not shown in the figure, are also expected to have a strong effect. These systematic errors include unaccounted-for attenuation in the source sample, contamination within mass, and TKE bins. An overall systematic correction of up to $20$\% on all points is expected due to efficiency estimation. Variation of this efficiency with energy introduces much smaller ($\approx 1$ \%) variation of this systematic correction across mass and TKE bins. While our results in that region are not statistically significant, a similar observation of multiplicity drop was reported by Gonnenwein \textit{et al.}~\cite{Gonnenwein2005} who interpreted it as an indication of the validity of the statistical model of AM generation. Their argument is that at the lowest TKE, \textit{i.e.}, the highest $E^*$ in the figure, the Coulomb potential is very weak, indicating a large separation of the fragments and thus a highly deformed configuration at scission. Because a lot of the energy is stored in nuclear deformations, it is not available to statistically excite the fragment rotational modes, thus leading to a drop in AM.

The results shown here are, to our knowledge, the first of their kind for $^{252}$Cf(sf), with the exception of Nifenecker \textit{et al.}~\cite{Nifenecker1972}, who reported a linear correlation between the total energy released by $\gamma$-rays and total neutron multiplicity, a proxy for $E^*$. In the same publication, the authors also determined a linear correlation between the $\gamma$-ray multiplicity and neutron multiplicity when gating on fragment masses. Nifenecker \textit{et al.} used these data to conclude that there exists a linear relationship between the fragment angular momenta and excitation energy. Our data, while different in the variables investigated, points to a different conclusion. A linearly increasing trend can be observed for some of the more symmetric and asymmetric fragmentations; however, this is not true overall for the fission reaction. On average, the relationship between total $\gamma$-ray multiplicity and $E^*$, or total neutron multiplicity, is not linear, but reaches a saturation plateau. 

\subsection{Individual $\gamma$-ray multiplicity}
\label{sub:indiv}

As explained in Section~\ref{sec:inst}, we can exploit the aberration of $\gamma$-rays in the forward and backward fragment propagation directions to separate the $\gamma$ ray multiplicities emitted by the light and heavy fragments. This technique, known as the Maier-Leibniz Doppler Shift analysis, has been optimized by Travar \textit{et al.}~\cite{Travar2021}, correcting the traditional techniques and reducing the effect of systematic biases. 

The drawback of this technique is that only two of the many detectors in FS-3 are used, significantly lowering the statistical significance of the experimental results. Each of the two detectors in this second configuration of the experiment can be analyzed independently of the other to determine the $\gamma$-ray multiplicity, thus reducing the effects of systematic error after averaging the two results. 

\begin{figure}[!htb]
\centering
\includegraphics[width=0.8 \linewidth]{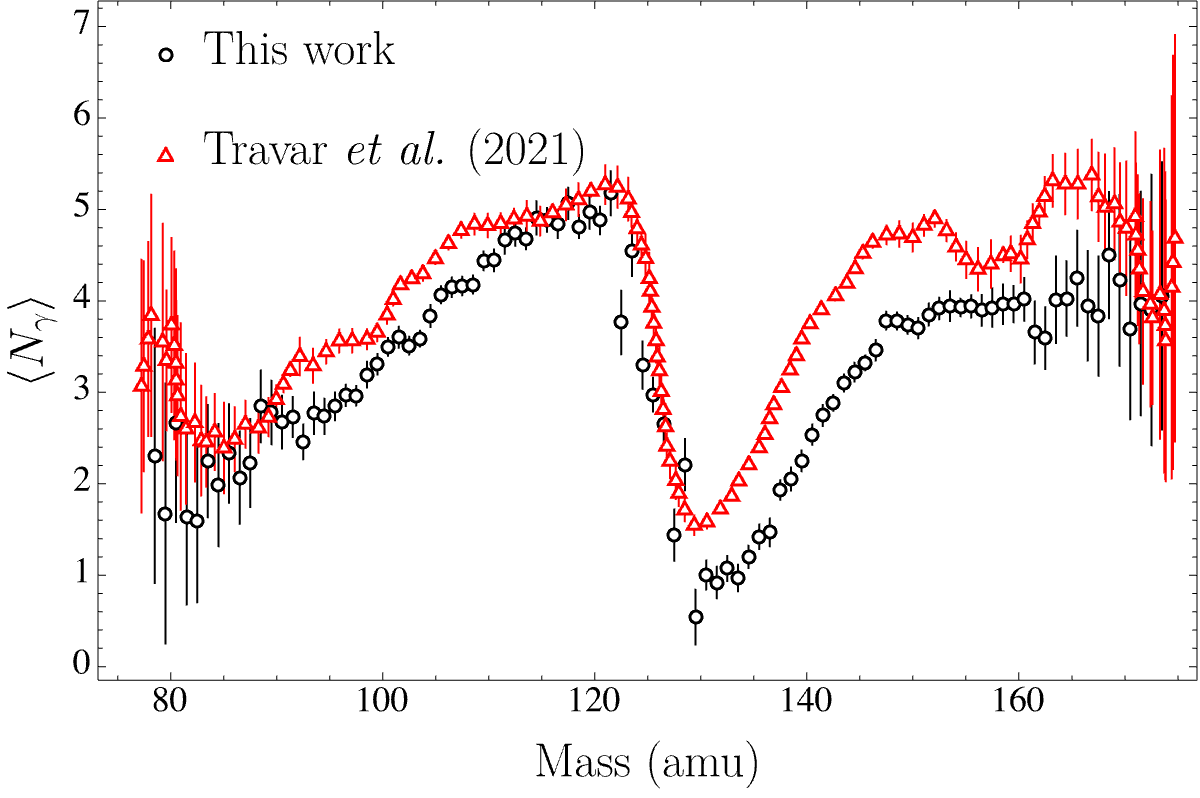}
\caption[Gamma sawtooth]{The dependence of the individual mean $\gamma$-ray multiplicity from each fragment as a function of their masses. The sawtooth shape is in good agreement with the investigation by Travar \textit{et al.}~\cite{Travar2021}. The results of our experiment are integral-normalized to be compared to the literature result. }
\label{fig:gamSaw}
\end{figure}

We show the individual mean $\gamma$-ray multiplicities as a function of fragment mass in Fig.~\ref{fig:gamSaw}. We find the results from both detectors to agree quite well with one another with respect to the mass-dependence. Our observations of a sawtooth pattern in the $\gamma$-ray multiplicity distribution align with recent findings by Travar \textit{et al.}~\cite{Travar2021}. These recent results, together with the direct measurement of the AM of several isotopes by Wilson \textit{et al.}~\cite{Wilson2021}, have confirmed the existence of an AM sawtooth in fission. 

In analogy to the neutron sawtooth~\cite{Brosa1990}, we interpret the $\gamma$-ray sawtooth, and its associated AM sawtooth, as indicative of a deformation-dependent AM. As already pointed out by Wilhelmy \textit{et al.}~\cite{Wilhelmy1972}, there is a strong correlation between the average AM of a fragment and the ground state electric quadrupole moment of the same isotope, a measure of its deformability. 

\begin{figure}[!htb]
\centering
\includegraphics[width=0.8 \linewidth]{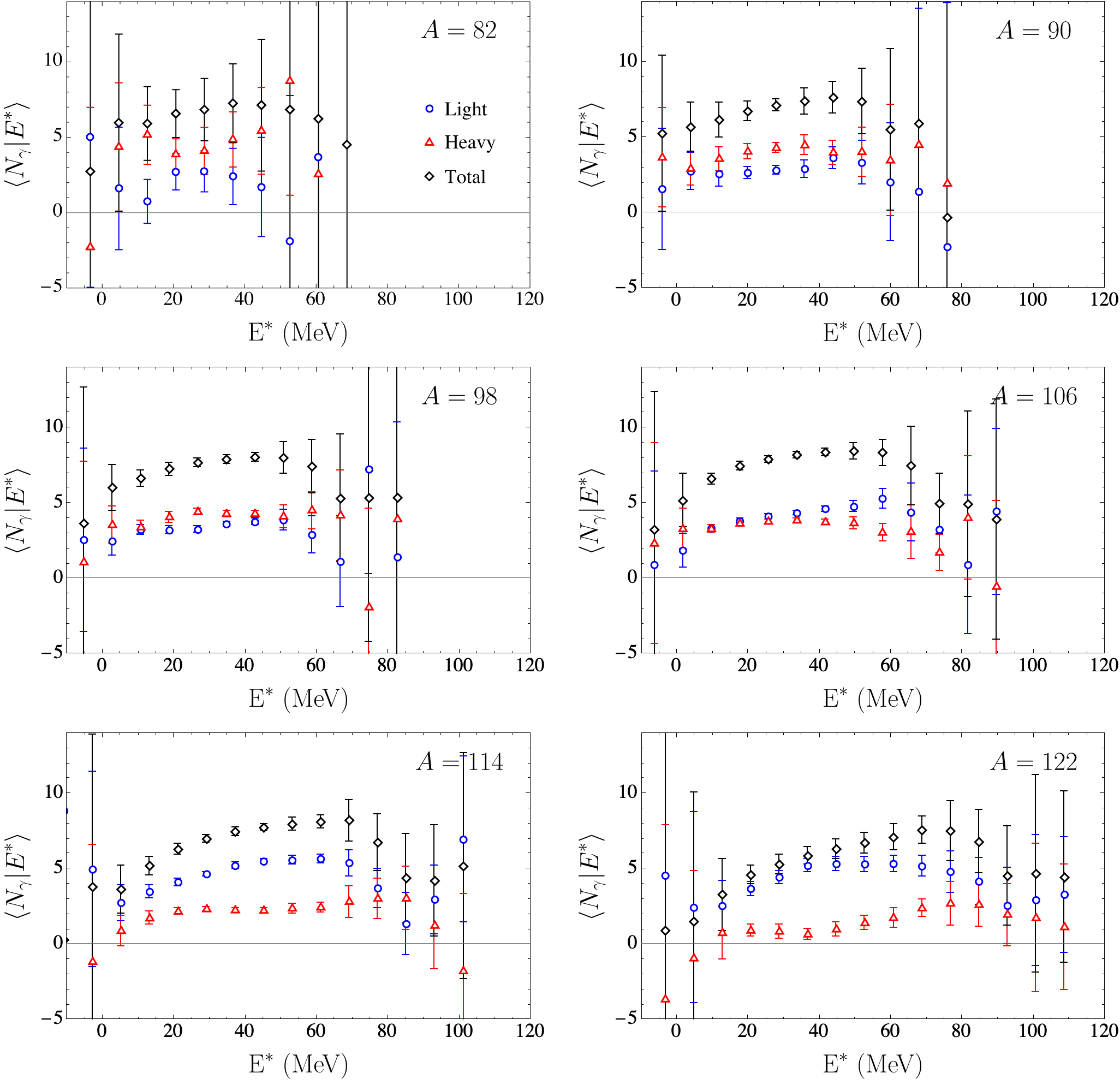}
\caption[Gamma mult individual]{Individual $\gamma$-ray multiplicity emitted from each fragment, binned by masses across $E^*$. Each panel is a separate fragmentation, with the bin center indicated on the figure and bin width of $8$ AMU. The sum of the two multiplicities is also shown. The energy broadening introduced by the detection system can lead to fragments having TKE above the total Q-value of the reaction; this effect results in the negative $E^*$ observed here.  }
\label{fig:gamIndiv}
\end{figure}

A new, insightful way of looking at the same data is to differentiate the $\gamma$-ray sawtooth with respect to TKE, as is done in Fig.~\ref{fig:gamIndiv}. Each panel in Fig.~\ref{fig:gamIndiv} shows a different mass split, with data points indicating the yield of $\gamma$ rays from the light and heavy fragment, as well as the total multiplicity. This plot shows the complicated relationship between the two fragments' $\gamma$-ray yield and, if interpreted as such, between the fragments' AM. We see that in general the $\gamma$-ray yield from the light and heavy fragment changes at different rates with $E^*$. For example, near the symmetric fragmentation, we see that the light fragment emits most of the $\gamma$ rays and only when this reaches a plateau does the heavy fragment begin to increase its $\gamma$-ray yield. For the most common fragmentation masses, $A_L \approx 110$, we find that the light fragment $\gamma$-ray yield increases with energy, while the heavy fragment yield remains constant if not slightly decreasing. This decrease in the $\gamma$-ray yield is not statistically significant given the current data, but is worth future exploration. In fact, a competition between the fragment AM magnitudes could signal the existence of excited bending modes in the statistical population of rotational modes. 

This result shows not only that $\gamma$-ray multiplicity increases at different rates for the light and heavy fragment, as one might expect from the presence of the $\gamma$-ray sawtooth, but also a competition between the $\gamma$ rays from opposite fragments. This effect can be seen for the more symmetric and asymmetric fragmentations, where we see the $\gamma$ ray multiplicity from one fragment increases while the multiplicity of the other fragment decreases with TKE. A similar result had been already observed by Pleasonton \textit{et al.}~\cite{Pleasonton1972} in thermal neutron-induced fission events. More recently, Piau \textit{et al.}~\cite{Piau2023} observed a similar trend by comparing the number of $\gamma$-rays emitted by the light and heavy fragment groups in general. These observations are at variance with the investigation of Wang \textit{et al.}~\cite{Wang2016}, which determined a more complicated relationship. In their investigation, Wang \textit{et al.} determined that the yield of $\gamma$ rays from the light fragment remains relatively constant with increasing excitation energy (as measured by the fragment's neutron multiplicity). For the heavy fragment, the $\gamma$-ray yield increases up to a maximum multiplicity, and then drops again. The observations of Wang \textit{et al.} do reproduce some of our own, for example the saturated and decreasing multiplicities, but cannot be compared directly since their mass bins are too large. One point of agreement is their observation that for symmetric fragmentation splits the $\gamma$-ray yield increases almost linearly with the excitation energy, as we have also observed. 

The roughly linear trend between $\gamma$-ray multiplicity and $E^*$ for the symmetric masses, previously discussed in connection to the results presented by Nifenecker \textit{et al.}, can now be understood better. In fact, we see that the linear trend arises as the sum of two approximately linear piecewise functions: at low $E^*$ the light fragment's $\gamma$-ray emission increases linearly with $E^*$ while the heavy fragment remains approximately constant; at higher $E^*$, the roles are reversed. When these two emissions are added together, the total emission appears to be linear over a wide range of $E^*$, but this trend also eventually saturates at the highest $E^*$. We also note that the symmetric region, especially in the vicinity of the 120-132 split that confers a high level of sphericity to the heavy fragment due to its closeness to the doubly magic nucleus $^{132}$Sn, the energy split between the two fragments is highly skewed towards the light fragment. In fact, based on the neutron sawtooth measurement, we can expect the light fragment to carry, on average, approximately 6-7 times more  energy than its heavy partner. Thus, the low $\gamma$-ray yield from the heavy fragment in this mass split is expected: the low excitation energy, coupled with the large level spacing in this mass region, reduces the population of high-spin states. 

\section{Angular distribution}
\label{sec:angGam}

Given the close connection between AM and rotational symmetry, the angular distribution of $\gamma$ rays often represents the most direct probe of AM in nuclear systems. To do this, we fit the experimental data using a sum of Legendre polynomials, 
\begin{equation}
    \sum_{i = 0}^{l_{\text{max}}} C_i P_i (\cos \theta)
\end{equation}
where $P_i$ is the i$^{th}$-order Legendre polynomial, and $C_i$ is the corresponding amplitude. The sum is taken up to order $l_{\text{max}}$, the choice of which does not influence the fit of the other components. In addition, to improve the fit further, each component in the fitting function is integrated within the experimental bins. Here, we examine how the anisotropy, defined as the ratio $C_2/C_0$ of Legendre polynomial coefficients, changes with the masses and TKE of the fission fragments. 


\subsection{$\gamma$-ray angular distribution}

The angular distribution of $\gamma$-rays is measured as the count rate in the detector in front of the TFGIC cylindrical axis. We then select those fission events that make an angle $\theta$ with this TFGIC axis. The procedure yields curves of $\gamma$-ray multiplicity over the angle $\theta$. As mentioned, we characterize the strength of the $\gamma$-ray anisotropy by taking the ratio of Legendre-polynomial-expansion coefficients. However, due to the limitations of the TFGIC and the fission target, angles with $|\text{cos}  \theta | < 0.3$ cannot be properly measured. We correct for this inaccessible region recursively, but we expect the Legendre coefficients so determined to have correlated systematic uncertainties.

\begin{figure}[!htb]
\centering
\includegraphics[width=0.8 \linewidth]{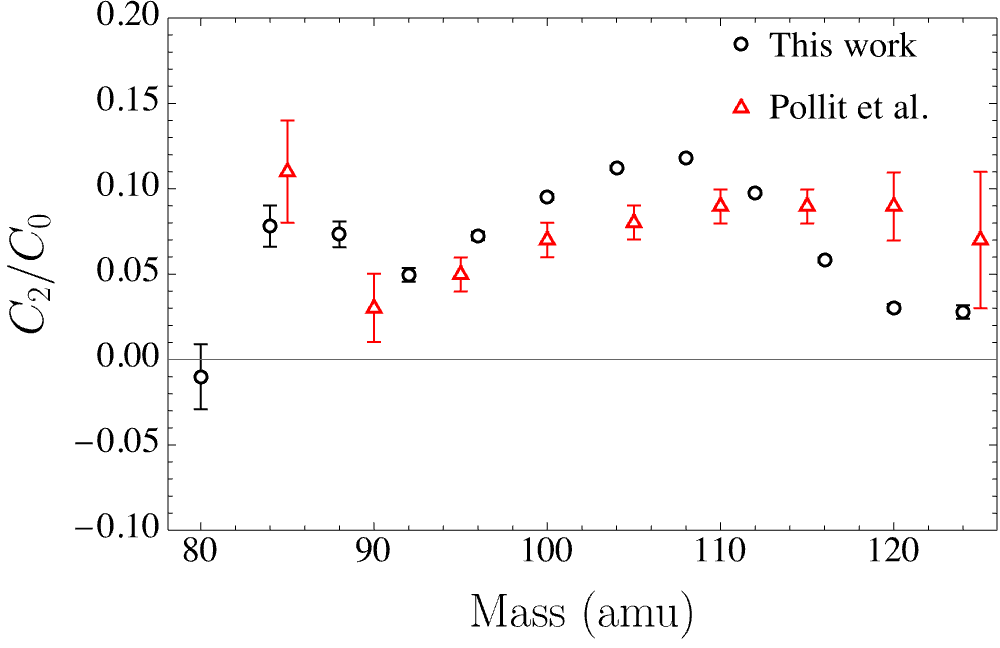}
\caption[Gamma angular distribution]{Mass-dependent anisotropy of the $\gamma$-ray angular distribution, as determined by the ratio of Legendre coefficients.}
\label{fig:gamAngDist}
\end{figure}

From physical arguments, we expect only even-order coefficient to have physical significance~\cite{Rose1953}. The odd-order coefficients can be induced by the aberration of $\gamma$ rays, a phenomenon exploited in the previous section for the separation of fragment emissions. In Fig.~\ref{fig:gamAngDist} we present the ratio $C_2/C_0$ as a function of the mass of the light fragment. 

It is a well-known fact in fission physics that the fragments' AM appear to be approximately aligned in a plane perpendicular to the fission axis~\cite{Wilhelmy1972, Hoffman1964}. Our results indicate a significant modulation of the $\gamma$-ray anisotropy with fragment mass. While not in quantitative agreement with the work by Pollitt \textit{et al.}~\cite{Pollitt2013}, especially in the region of symmetric split, both results agree on the levels of variation in the anisotropy across the mass spectrum. The changing anisotropy could be an indication of a different degree of the polarization of the fragments' AM; however, this interpretation is not yet warranted by these data on their own. Several confounding variables could be participating in this phenomenon, including a depolarization of the AM by neutron evaporation and statistical $\gamma$-ray emission. 

We note that the highest level of anisotropy, suggestive of a high degree of AM polarization, is found at light-fragment masses near 110. On the other hand, the lowest anisotropies, indicating a more isotropic AM distribution with respect to the fission axis, are found at the very symmetric and asymmetric extremes. The increase in anisotropy at approximately $A_L \approx 84$ is in agreement with the previously observed feature in Pollitt \textit{et al.}~\cite{Pollitt2013}. 

\subsection{Channel differentiation}

\begin{figure}[!htb]
\centering
\includegraphics[width=0.55 \linewidth]{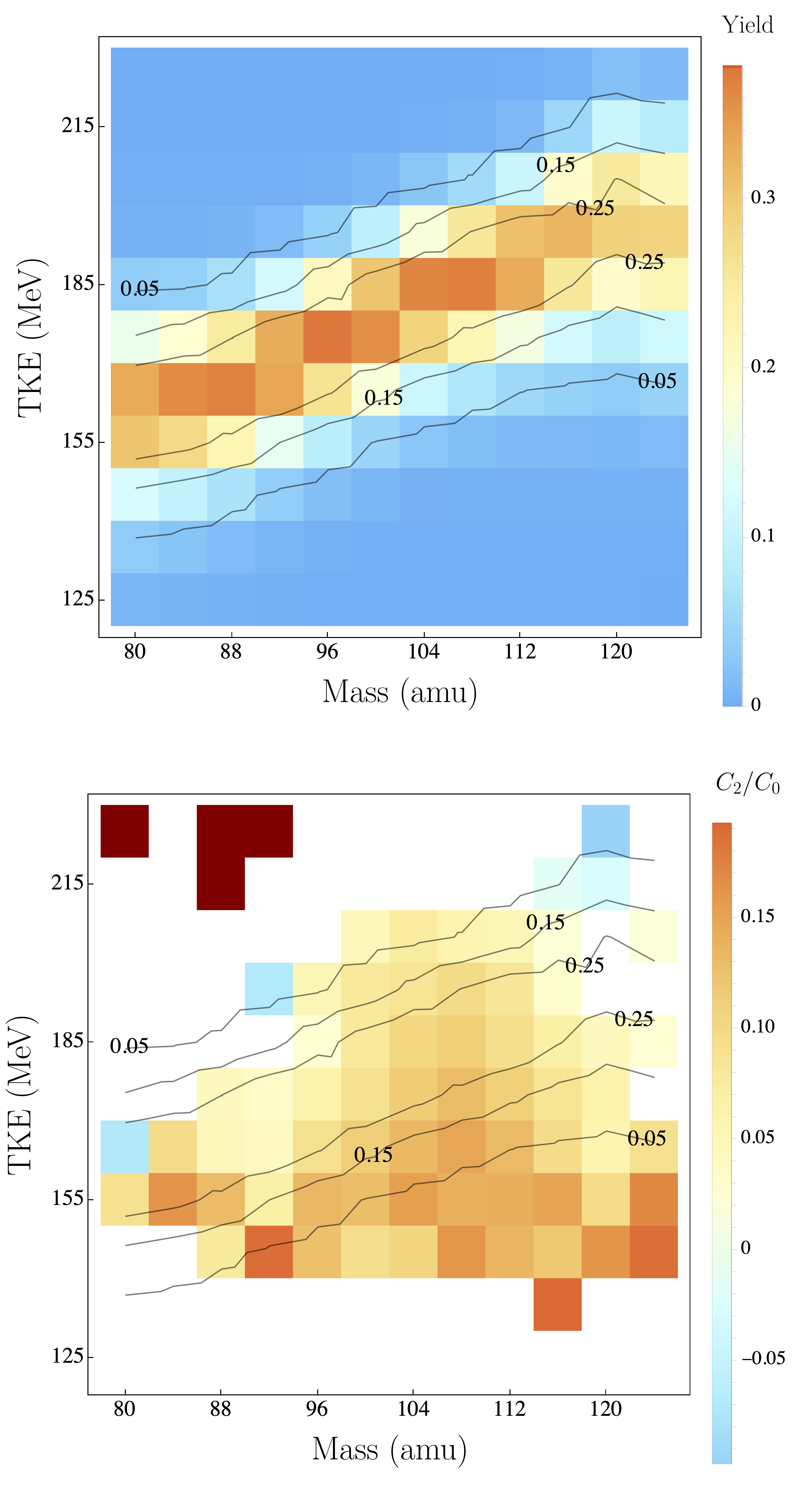}
\caption[Grid ang]{Yield-differentiated $\gamma$-ray anisotropy coefficients. The top panel shows the fragment yields themselves. The bottom panel shows the anisotropy of $\gamma$ rays for the different mass regions. Contour lines are determined from the yield. Data with $> 30$ \% statistical uncertainty has been removed.}
\label{fig:grAng}
\end{figure}

The data presented in Fig.~\ref{fig:gamAngDist} can be further differentiated with respect to TKE. This differentiation allows us to gain an insight into the dynamics of the fission process itself. Yields of mass and TKE have often been employed in the literature to draw conclusions about scission shapes and configurations. The discussion of standard, symmetric, and asymmetric splits should be familiar from the classic description of fission as a multi-channel process~\cite{Brosa1990}. According to this formalism, fission can proceed through several reaction channels, each characterized by a shape of the fissioning nucleus at the moment of fission. These channels differentiate, for example, between spherical and elongated shapes, associated with high and low TKE due to the different strength of the Coulomb repulsion; and between symmetric and asymmetric mass splits, as already pointed out earlier. The result of this further differentiation of the $\gamma$-ray angular anisotropy data is presented in Fig.~\ref{fig:grAng}. The anisotropy of $\gamma$ rays, expressed by  the ratio $C_2/C_0$, is more pronounced at lower TKE,  and there appears to be especially strong anisotropies at symmetric masses.

In the first place we note that, in general, the anisotropy increases when TKE decreases. This result is reminiscent of the multiplicity saturation discussion presented in the previous section. Our anisotropy results also show a similar mass dependence, showing that at sufficiently high excitation energies, or conversely low TKE, an anisotropy in the $\gamma$-ray yield can be observed. Unfortunately the amount of data collected is not sufficient to draw quantitative conclusions across the entire yield.

We recognize the four fission channels~\cite{Brosa1986} in our figure, a standard channel around $A_L = 110$ and TKE$~\sim180$ MeV; super-long and super-short channels at symmetric masses $A_L \sim 125$ and TKE$~\sim140$ and TKE$~\sim210$, respectively; a super-asymmetric channel at $A_L \sim 85$ and TKE$~\sim140$. Most noticeable is the presence of strong polarization in the regions we associate with the standard, super-long, and super-asymmetric channels. These channels are associated with the larger degree of deformation in the scissioning configuration, thus suggesting an interpretation that the magnitude and direction of the fragments' AM are closely connected to the shape of the nucleus at scission. On the other hand, we do not observe a significant degree of polarization in the super-short channel, which is characterized by the least deformed configuration. These observations give credit to a scission deformation-dependent AM generation mechanism.

\section{Conclusion}
\label{sec:conc}

We have presented new experimental results on $^{252}$Cf(sf) from a recent experiment performed using a combination of a TFGIC and organic scintillators. The results add important information to the already rich literature on nuclear fission and specifically on the generation of AM in fission. Our three main conclusions stem from our results regarding the multiplicity and the angular distribution of $\gamma$ rays: saturation, competition, and polarization. 

We have determined that the multiplicity of $\gamma$ rays does not increase steadily with fission fragment excitation energy. On the contrary, we have determined that the mean $\gamma$-ray multiplicity increases up to a value of approximately $8$ $\gamma$ rays and appears to level off after that. This could be an indication of a saturation of the AM, which would significantly restrict the ways in which AM is generated, but could also be explained by a larger portion of AM being carried by neutrons and statistical $\gamma$ rays. Future experiments should focus on the issue of AM removal by this type of radiation. If this saturating pattern persists, it would indicate the existence of non-statistical mechanisms in the generation of AM. In combination with the existence of the AM sawtooth, this result indicates the importance of deformation-dependent mechanisms of AM generation. 

By exploiting the aberration of $\gamma$ rays emitted by moving sources, we have separated the average emission of $\gamma$ rays for the light and heavy fragment. The results of this analysis indicate that the emission from the two fragments do not necessarily grow together with $E^*$. Instead, we observe that the emission can be uncorrelated, with the increase of $\gamma$ rays being produced primarily by one or the other fragment, or even anti-correlated, where the emission of $\gamma$ rays increases in one fragment while decreasing in the other. If these findings are indeed indicative of AM patterns and trends, they would indicate the presence of complex modes of AM generation.  

Lastly, we have challenged an important and established experimental observation in fission, namely the predominance of transversely polarized fragment AM. Specifically, we observed that the $\gamma$-ray anisotropy possesses a mass and TKE dependence. The strongest anisotropy, corresponding to the most strongly transversely polarized fragments' AM, were found in correlation with the super-long channel, the channel with the most deformation at scission. The polarization of AM could be a determining factor in the development of future models of AM generation in fission. 

Overall, the experimental results of this paper challenge some of the assumptions made in the discussion of $\gamma$-rays in fission, especially regarding the mass and TKE dependence of their multiplicity and angular distribution. Future research should be focused on understanding the impact of the emission of neutrons and statistical $\gamma$ rays on the fragment AM, as the study of the rotational band $\gamma$ rays has remained the most prominent technique for AM assignment. 

\acknowledgments~\\[-4ex]

The authors would like to thank Olivier Litaize and the developers of the \texttt{FIFRELIN} code for providing the analyzed events as well as useful discussion. This work was in part supported by the Office of Defense Nuclear Nonproliferation Research and Development (DNN R \& D), National Nuclear Security Administration, US Department of Energy. This work was funded in-part by the Consortium for Monitoring, Technology, and Verification under Department of Energy National Nuclear Security Administration award number DE-NA0003920. This material is based upon work supported by the U.S. Department of Energy, Office of Science, Office of Nuclear Physics, under Contract Number DE-AC02-06CH11357. This research used resources of Argonne National Laboratory's ATLAS facility, which is a DOE Office of Science User Facility. Authors also gratefully acknowledge the use of the Bebop cluster in the Laboratory Computing Resource Center (LCRC) at Argonne National Laboratory. The work of V.A.P.  was performed under the auspices of UT-Battelle, LLC under Contract No. DE-AC05-00OR22725 with the U.S. Department of Energy. SM is grateful to Prof. Kiedrowski at the University of Michigan for helpful discussion regarding Legendre polynomial extraction. Authors would like to acknowledge the efforts in the design and support during operation provided by the LETS group at the Physics Division of ANL, especially M. Oberling and R. Knaack.

\bibliography{mybib}

\end{document}